\documentclass{aa} 
\usepackage{natbib,graphicx}
\begin{document}
\title{Astrometric and Spectroscopic Confirmation of a Brown Dwarf Companion to GSC\,08047-00232\thanks{Based on observations obtained at the VLT in ESO with programs 70.C-0677 and 072.C-0644} }
\subtitle{VLT/NACO deep imaging and spectroscopic observations}
\titlerunning{VLT/NACO AO imaging and spectroscopy of GSC\,8047-0232}
\authorrunning{Chauvin et al.}
\author{
        G. Chauvin\inst{1}\and
        A.-M. Lagrange\inst{2}\and
	F. Lacombe\inst{3}\and
        C. Dumas\inst{1}\and
        D. Mouillet\inst{4}\and
        B. Zuckerman\inst{5}\and
	E. Gendron\inst{2}\and
	I.~Song\inst{5}\and
        J.-L. Beuzit\inst{1}\and
        P. Lowrance\inst{6}\and
	T. Fusco\inst{7}
}
\offprints{Ga\"el Chauvin \email{gchauvin@eso.org}}
\institute{
$^{1}$European Southern Observatory, Casilla 19001, Santiago 19, Chile\\
$^{2}$Laboratoire d'Astrophysique, Observatoire de Grenoble, 414, Rue de la piscine, Saint-Martin d'H\`eres, France\\
$^{3}$Laboratoire D'Etudes Spatiales et d'Instrumentation en Astrophysique, Observatoire de Paris, Bat. 16, 5 Place J. Jansen, 92195 Meudon, France\\ 
$^{4}$Laboratoire d'Astrophysique, Observatoire Midi-Pyr\'en\'ees, Tarbes, France\\
$^{5}$Department of Physics \& Astronomy and Center for Astrobiology, Physics \&
Astronomy Building, UCLA, Los Angeles, CA 90095-1562, USA\\
$^{6}$SIRTF Science Center Infrared Processing and Analysis Center, MS 220-6, Pasadena, CA 91125, USA\\
$^{7}$ONERA, BP52, 29 Avenue de la Division Leclerc Ch\^atillon Cedex, France \\
}
\date{Received:  / Accepted: }
\abstract{We report VLT/NACO imaging observations of the stars GSC\,08047-00232 and HIP\,6856, probable members of the large Tucana-Horologium association. During our previous ADONIS/SHARPII deep imaging survey, a substellar candidate companion was discovered around each star. Based on VLT/NACO astrometric measurements, we find that GSC\,08047-00232 and the faint candidate companion near to it share the same proper motion with a significance of 3.1\,$\sigma$. On the contrary, the candidate companion to HIP\,6856 is probably a background object with a significance of 4.3\,$\sigma$. We also detect a new fainter and closer candidate companion to HIP\,6856, but which is likely a background object too with a significance of 4\,$\sigma$. Recent VLT/NACO spectroscopic measurements of GSC\,08047-00232 finally confirm the substellar nature of this young brown dwarf with a derived spectral type M$9.5\pm1$. GSC\,08047-00232\,B, with an estimated mass of $25\pm10\,$~M$_{\rm{Jup}}$ and an effective temperature of $2100\pm200$\,K, is presently the third substellar companion identified among young, nearby associations. 
\keywords{instrumentation: adaptive optics  --- stars: imaging --- stars: low-mass, brown dwarfs}}
\maketitle
%
%
\section{Introduction}

Since the discovery of the first bound substellar objects GD\,165~B \citep{beck88} and Gl\,229~B \citep{naka95}, deep IR surveys have revealed the existence of numerous isolated brown dwarfs (BDs) \citep{mora03}. These surveys reveal that BDs at large separations (typically $\geq 1000$\,AU) may be common as companions to stars \citep{gizi01}. At closer separations ($\leq4\,$AU), radial velocity surveys are detecting an increasing number of planets around nearby stars and have indicated a lack of BD companions, known as the BD desert \citep{halb00}. 
\begin{table*}[!t]
\caption[]{NACO observations: instrumental set-up and observations logs. Camera S27 and S13 respectively provide ~27.0 and 13.24 mas/pixel sampling. DIT and NDIT correspond respectively to the exposure integration time and the number of exposures.}
\centering
\begin{tabular}{lllllllll}
\hline\hline\noalign{\smallskip}
Star 		& UT Date      & NDIT$\times$DIT & Filter         & Camera   & Mode       & Seeing   & Remarks \\
                &  	       &                 &                &          &            & ($\!``$)     &      	  \\

\noalign{\smallskip}\hline\noalign{\smallskip}
\multicolumn{8}{c}{Classical and Coronagraphic Imaging}\\
\noalign{\smallskip}\hline\noalign{\smallskip}
GSC\,08047-00232\,AB  & 23/11/2002  & 100$\times$3s   & NB1.24        & S13 & classical          &   0.82   & science        \\
GSC\,08047-00232\,AB  & 23/11/2002  & 100$\times$3s   & NB1.64        & S13 & classical          &   0.99   & science        \\
GSC\,08047-00232\,AB  & 23/11/2002  & 100$\times$0.6s & NB2.17 	 & S27 & classical          &   0.78   & science       \\
GSC\,08047-00232\,AB  & 23/11/2002  & 100$\times$3s   & J 		 & S13 & saturated        &  0.65         & science   \\
GSC\,08047-00232\,AB  & 23/11/2002  & 100$\times$2s   & H 		 & S13 & saturated        &  1.05         & science   \\
GSC\,08047-00232\,AB  & 23/11/2002  & 100$\times$1s   & Ks 		 & S27 & saturated        &  0.69         & science   \\
HIP\,6856       & 26/11/2002  & 20$\times$4s  & J+ND1$^{a}$     & S13 & classical          &  0.75          & science          \\
HIP\,6856       & 26/11/2002  & 3$\times$60s   & J               & S13 & coronagraphic    &  0.72          & science          \\
HIP\,6856       & 26/11/2002  & 30$\times$0.34s & NB2.17          & S27 & classical         &  0.57          & science          \\
HIP\,6856       & 26/11/2002  & 10$\times$15s   & Ks              & S27 & coronagraphic   &  0.76          & science          \\
HD\,7644        & 23/11/2002  & $\sim$60s       & Each    & Each     & classical &        0.7-0.9       & standard phot.  \\
Orion\,n           & 21/11/2002  & 100$\times$2s   & H              & S13 & classical   &  1.1-1.2             & standard astro.      \\  
\noalign{\smallskip}\noalign{\smallskip}
HIP\,6856       & 08/06/2003  & 5$\times$4s               & J+ND1$^{a}$               & S13 & classical          &  0.85           & science          \\
HIP\,6856       & 08/06/2003  & 2$\times$60s               & J               & S13 & coronagraphic    &  0.61           & science          \\
HD\,114842        & 29/05/2003  & 40$\times$0.5s               & NB1.64               & S13 & classical          &  0.82       & standard astro.      \\
HD\,208372        & 30/05/2003  & 40$\times$2s               & NB1.64               & S13 & classical          &  0.39       & standard astro.      \\
\noalign{\smallskip}\noalign{\smallskip}
HIP\,6856       & 07/09/2003  & 5$\times$3s   & J+ND1$^{a}$     & S13 & classical          &  0.54       & science          \\
HIP\,6856       & 07/09/2003  & 6$\times$60s   & J               & S13 & coronagraphic    &  0.56       & science          \\
GSC\,08047-00232\,AB  & 07/09/2003  & 10 $\times$2s & NB1.64          & S13 & classical          &  0.6        & science          \\
GSC\,08047-00232\,AB  & 07/09/2003  & 80$\times$2.5s & H               & S13 & saturated        &  0.75       & science          \\
HD\,211742        & 07/09/2003  & 5$\times$3s               & NB1.64               & S13 & classical          &  0.82       & standard astro.      \\
\noalign{\smallskip}\noalign{\smallskip}
GSC\,08047-00232\,AB  & 05/03/2004  & 4$\times$3s & NB1.64          & S13 & classical          &  0.84        & science          \\
GSC\,08047-00232\,AB  & 05/03/2004  & 4$\times$8s & H               & S13 & saturated        &  1.12       & science          \\
$\theta$\,Ori           & 05/03/2004  & 10$\times$12s   & NB2.17              & S13 & coronographic   &  1.3-1.4             & standard astro.      \\  

\noalign{\smallskip}\hline\noalign{\smallskip}
\multicolumn{8}{c}{Spectroscopy}\\
\noalign{\smallskip}\hline\noalign{\smallskip}
GSC\,08047-00232\,B  & 25/11/2002  & 3$\times$600s               & SK-Grism        & S54 & R=1400        &  0.82       & science          \\
HIP\,9022     & 25/11/2002  & 12$\times$1s               & SK-Grism        & S54 & R=1400        &  0.68       & standard tell.         \\
\noalign{\smallskip}
\hline
\end{tabular}
\label{tab:setup}
\begin{list}{}{}
\item[$^{\mathrm{a}}$] ND1 is a CONICA neutral density filter with a transmission of 1.4\%
\end{list}
\end{table*}

The intermediate separation range between 4 and 1000~AU is just beginning to be explored as, from the ground, large ground-based telescopes coupled with adaptive optics (AO) and coronographic instruments are required. Surveying G, K and M-type nearby ($\leq25$~pc) stars with a typical age of $\sim300$~Myr, \citet{mcar04} recently showed that the brown dwarf desert extends from 75 to 1200~AU. 

Exploring a comparable separation range with the instrument ADONIS/SHARPII at the ESO/3.6m telescope, we conducted on November 2000 and October 2001 an AO coronagraphic imaging survey of the large common Tucana-Horologium association (see Chauvin et al., 2003). This large 30~Myr old association has been previously identified by \citet{torr00} and \citet{zuck00} with a distance estimated at $\sim50$\,pc from Earth. Based on the ADONIS/SHARPII observations of two dozen probable association members, we reported the detection of one substellar candidate companion (hereafter cc-1) to the star HIP\,6856 (K1V, V=9.39, $d=37.1_{-1.6}^{+1.8}~$pc) and another one to GSC\,08047-00232 (K3V, $V=10.9$). The faint object close to GSC\,08047-00232 was independently detected by \citet{neuh03} with the SHARP instrument at the ESO \textit{New Technology Telescope} (NTT). Recently, Neuh\"auser \& Guenther (2004) acquired H- and K-band spectra of this faint object and derived a spectral type M$8\pm2$, but could not confirm its companionship without proper motion measurements. 

Although GSC\,08047-00232 is undoubtedly a young star ($\rm{W}_{\rm{Li}}=350~\rm{m}\AA$) and certainly a probable member of the Tucana-Horologium association, this star was not included among the latest volume-limited ($\textit{d}\leq60$~pc) selection of Zuckerman \& Song (2004) in order to limit their sample size. Although no trigonometric parallaxes are known for GSC\,08047-00232, a distance of 85\,pc (adopted hereafter) can be derived according to the JHK photometry of the 2MASS All-Sky Catalog of Point Sources \citep{cutr03}. This is supported by an independent distance estimation of 87~pc derived from a $\rm{V}-\rm{K}$ versus $\rm{M}_{K}$ diagram (Song et al. 2004, in prep). 
\begin{figure*}[t]
\centering
\includegraphics[width=15.cm]{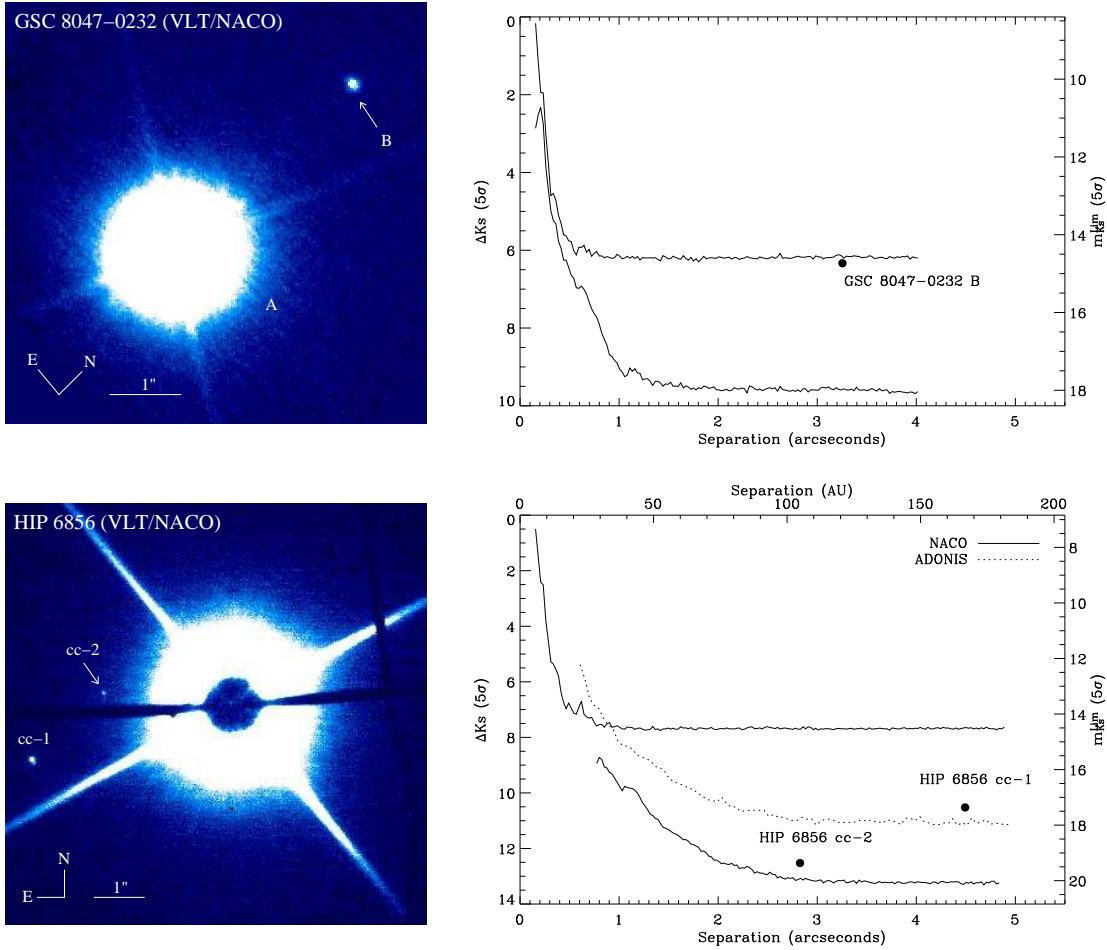}
\caption{\textbf{Top}: Left, NACO image of GSC 08047-00232 A and B observed in Ks with the S27 camera. Right, NACO (pixel to pixel) detection limits of GSC~08047-00232 obtained in classical imaging in NB2.17 ($100\times0.6$s, \textit{upper solid line}) and in saturated imaging Ks ($100\times1$s, \textit{lower solid line}). \textbf{Bottom}: Left, Coronagraphic image of the source HIP\,6856 obtained in Ks with an occulting mask of diameter $1.4~\!''$ and the S27 camera. The candidate companion cc-1 ($\Delta\rm{Ks}=10.6$), imaged with ADONIS/SHARPII \citep{chau03}, and the new candidate cc-2 ($\Delta\rm{Ks}=12.8$), detected with VLT/NACO, are indicated. Right, NACO (pixel to pixel) detection limits of HIP\,6856 obtained in classical imaging mode in NB2.17 ($5\times0.34$s, \textit{upper solid line}) and in coronagraphy in Ks ($10\times15$s, \textit{lower solid line}). The detection limit obtained with the coronagraphic imaging mode of ADONIS/SHARPII during $15 \times 20$s on 29 October 2001 is also reported (\textit{dotted line}).}
\label{fig:ao}
\end{figure*}

To study the proper motions of these faint objects, new observations were obtained with NACO at the VLT/UT4 at different epochs. NACO enabled us to detect a second closer and fainter candidate companion (hereafter cc-2) to the star HIP\,6856. 

We present a brief description of the observational and data reduction techniques. Then, we discuss the status of the three candidate companions based on: 1) the consistency of their photometry with the predictions obtained from evolutionary models for very low mass objects, 2) the study of their proper motion in comparison with their primary star, 3) the acquisition of a near-IR spectrum to corroborate the companionship and to confirm the substellar nature. The results of this analysis, given in sections 3, 4 and 5, lead to the rejection of the two candidate companions as bound objects to HIP\,6856 and the discovery of a young late-M brown dwarf companion (which we designate as B) to the star GSC\,08047-00232\,A.

%
%
%
\section{Observations and data reduction}

HIP\,6856 and GSC\,08047-00232 were imaged with the NAOS/CONICA (NACO) instrument of the VLT-UT4. The AO system NAOS \citep{rous02} is equipped with a tip-tilt mirror, a 185 piezo actuator deformable mirror and two wavefront sensors (Visible and IR). Attached to NAOS, CONICA \citep{lenz98} is the near infrared ($1-5\,\mu$m domain) imaging, Lyot coronagraphic, spectroscopic and polarimetric camera, equipped with a $1024\times1024$ pixels Aladdin InSb array. 

\subsection{Classical and coronagraphic imaging}

To image the faint candidate companions to HIP\,6856, the coronagraphic mode was used with an occulting mask of diameter $1.4~\!''$ and with the J and Ks filters. 

In the case of the GSC\,08047-00232, the narrow band filters NB1.24, NB1.64 and NB2.17 were used in classical imaging mode to avoid saturation of the primary. To image the faint candidate companion, we used the J, H and Ks filters, integrating the primary above the 1\% linearity level of the CONICA detector (i.e in saturated imaging). The relative position between the candidate companion and the primary could then be determined by taking into account the shifts between narrow band and broad band filters. The instrumental set-up and the observation logs are reported in Table~\ref{tab:setup}. The detection performances achieved with VLT/NACO for these two objects are reported in Fig.~\ref{fig:ao}. 

Determining companionship through proper motion of the candidate companions detected with ADONIS/SHARPII was the main interest in re-imaging GSC\,08047-00232 and HIP\,6856. As the cc-2 to HIP\,6856 was discovered thanks to NACO's detection capability, aperture photometry was used to obtain the J and K photometry of this faint object. For each epoch, the relative positions of the cc-1 and cc-2 to HIP\,6856 were then obtained with \textit{IRAF/DAOPHOT}, determining the photo-centers by a Gaussian centering. In the case of GSC\,08047-00232 A and B, the deconvolution algorithm of \citet{vera98} was used to determine the photometry of both components but also to estimate their relative positions at each epoch. For flux calibration, the IR photometric standard HD~7644 \citep{vand96} was observed on November 2002 in each observing set-up used for the science objects. The photometry of the candidate companions and their primaries is reported in Table~\ref{tab:photo}. The photometric measurements from the 2MASS All-Sky Catalog of Point Sources \citep{cutr03} are also given and demonstrate the good fidelity of the NACO photometry in the case of GSC\,08047-00232\,A. 

Astrometric calibrations were obtained at each epoch, by observing either the sources Orion\,n and $\theta$ Ori from the OMC-1 complex \citep{luhm00} respectively on 22 November 2002 and 5 March 2004, the astrometric references HD\,114842 and HD\,208372 on 29-30 May 2003, and HD\,211742 on 7 September 2003. Instrumental settings and observation logs for these astrometric standards are also reported in Table~\ref{tab:setup}. To obtain the best astrometric precision for our proper motion measurements, the S13 camera was used and calibrated. The orientations of true north of the S13 camera were found on 21 November 2002, 29-30 May 2003, 7 September 2003 and 5 March 2004, respectively at $-0.05^o,-0.1^o, +0.05^o,+0.04^o$ east of the vertical within an uncertainty of $0.1^o$. The pixel scale was found stable in time with a value of $13.24\pm 0.05$\,mas. The relative positions of the candidate companions and their primaries for different epochs are reported in Table~\ref{tab:astro}. The uncertainties take into account errors of the position estimation on the detector of both components as well as errors of the pixel scale and the orientation of the detector. 

We did not consider here the astrometric positions determined with ADONIS/SHARPII as their uncertainty was large in comparison to the precision achieved with VLT/NACO. \citet{neuh03} already determined the relative position of GSC\,08047-00232\,A and B on 4 July 2001 using speckle imaging with SHARP at the ESO/NTT telecope. However, this measurement is not consistent (at less than $3\sigma$) with the VLT/NACO relative positions that we obtained at three different epochs regarding the proper motion of GSC\,08047-00232\,A. Therefore, we only considered here the VLT/NACO observations to test if GSC\,08047-00232\,B is co-moving with GSC\,08047-00232\,A.

\subsection{Spectroscopy}

GSC\,08047-00232\,B was also observed in spectroscopy with NACO on 25 November 2002. The medium resolution ($\rm{R}_{\lambda}=1400$) grism was used with the 86~mas slit, the S54 camera (54~mas/pixel) and the SK filter (1.79-2.42~$\mu$m). The standard star HIP\,9022 (B8/B9V) was also observed to remove the telluric lines. After substracting the sky and dividing by a flat field using \textit{eclipse} \citep{devi97}, the spectra of GSC 08047-00232 B and HIP\,9022 were extracted and calibrated in wavelength with \textit{IRAF/DOSLIT}. To calibrate the relative throughput of the atmosphere and the instrument, we divided the extracted spectra of GSC 08047-00232 B by the spectra of HIP\,9022 and then multiplied by a blackbody to restore the shape of the continuum. GSC\,08047-00232\,B, separated by about $3.2~\!''$, was not contaminated by GSC\,08047-00232\,A.  

%
%
%
\section{Photometry and evolutionary models} 

\begin{table}[b]
\centering
\caption[]{Photometry of GSC\,08047-00232 A and B (designated as GSC\,08047 A and B below) and of the cc-1 and cc-2 to HIP\,6856}
\begin{tabular}{llll}
\hline\hline\noalign{\smallskip}
Source		&J		&H                 &K                  \\
		&(mag)		&(mag)             &(mag)	        \\
\noalign{\smallskip}\hline\noalign{\smallskip}

\scriptsize{HIP\,6856}$^{a}$ 	                &$7.41\pm0.02$	&$6.94\pm0.03$     &$6.83\pm0.03$       \\
\scriptsize{HIP\,6856}\,cc-1$^{b}$	        &$17.97\pm0.20$	&                  &$17.40\pm0.15$      \\
\scriptsize{HIP\,6856}\,cc-2$^{c}$	        &$19.9\pm0.4$	&                  &$19.6\pm0.2$   	\\
\noalign{\smallskip}\hline\noalign{\smallskip}
\scriptsize{GSC\,08047}\,A$^{a}$ &$9.06\pm0.03$  &$8.53\pm0.06$	   &$8.41\pm0.03$	\\
\scriptsize{GSC\,08047}\,A$^{c}$ &$9.06\pm0.09$  &$8.54\pm0.05$	   &$8.45\pm0.07$	\\
\scriptsize{GSC\,08047}\,B$^{c}$  &$15.90\pm0.13$ &15.45$\pm$0.20	   &$14.75\pm0.13$	\\
\noalign{\smallskip}\hline
\end{tabular}\label{tab:photo}
\begin{list}{}{}
\item[$^{\mathrm{a}}$] from the 2MASS All-Sky Catalog of Point Sources \citep{cutr03}.
\item[$^{\mathrm{b}}$] from \citet{chau03}.
\item[$^{\mathrm{c}}$] NACO measurements presented in this work.
\end{list}
\end{table}

\begin{table*}[t]
\centering
\small{
\caption[]{Relative positions of the cc-1 and cc-2 to HIP\,6856, and of GSC\,08047-00232\,B to GSC\,08047-00232\,A . }
\label{tab:astro}
\begin{tabular}{lllllllll}
\hline\hline\noalign{\smallskip}
Source		 &UT Date 	&Sep 		&P.A. 		& $\Delta$Sep 	& $\Delta$P.A. \\
                 	 	 &        &(mas)          & ($^o$)	& (mas) 	& ($^o$)	\\
\noalign{\smallskip}\hline\noalign{\smallskip}
HIP\,6856 cc-1	& 26/11/2002	&$4585\pm14$	&$105.7\pm0.3$ 	&-	        &-	\\
		&08/06/2003	&$4547\pm14$	&$105.8\pm0.3$	&$-38\pm20$	&$+0.1\pm0.4$	\\
		&07/09/2003	&$4503\pm13$	&$105.4\pm0.3$ 	&$-82\pm19$ 	&$-0.3\pm0.4$ 	\\
\noalign{\smallskip}
HIP\,6856 cc-2	&26/11/2002	&$2830\pm13$	&$85.4\pm0.3$ 	& -		& -  		\\
		&08/06/2003	&$2791\pm15$	&$85.3\pm0.3$	&$-39\pm20$ 	&$-0.1\pm0.4$ 	\\
		&07/09/2003	&$2749\pm14$	&$85.2\pm0.3$	&$-81\pm19$ 	&$-0.2\pm0.4$ 	\\
\noalign{\smallskip}\hline\noalign{\smallskip}
GSC\,08047-00232 B &23/11/2002  	&$3274\pm12$	&$358.85\pm0.23$&-	        &-	\\
		 &07/09/2003  	&$3266\pm11$	&$358.89\pm0.23$&$-8\pm16$      &$0.04\pm0.31$\\
		 &05/03/2004  	&$3260\pm12$	&$358.82\pm0.22$&$-14\pm15$     &$0.02\pm0.30$\\
\noalign{\smallskip}\hline
\end{tabular}
}
\end{table*}
\begin{figure}[t]
\centering
\includegraphics[width=8.5cm]{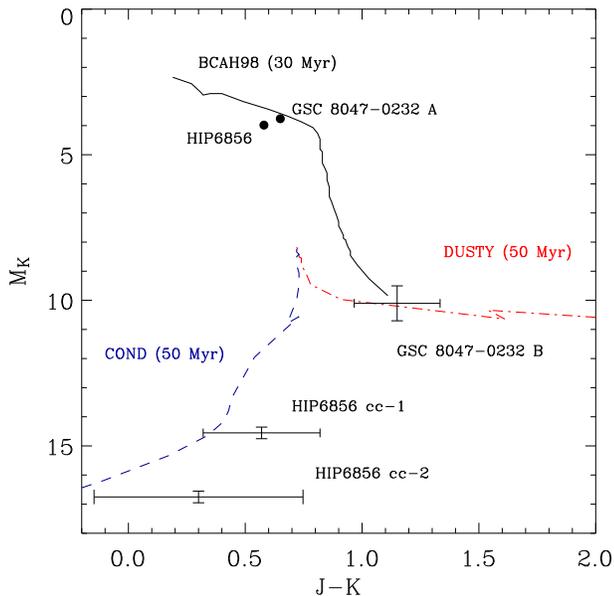}
\caption{Color-magnitude diagram for the near IR color $\rm{J}-\rm{K}$ from the BCAH98 model for an age of 30~Myr and for the CBAH00 DUSTY and COND models for an age of 50~Myr (no predictions at 30~Myr for these two models). The data of the cc-1 and cc-2 to HIP\,6856 as well as GSC\,08047-00232\,B are overplotted with their corresponding uncertainties. The two black points represent the two primaries HIP\,6856 and GSC\,08047-00232\,A. As discussed previously, a distance of 85~pc has been considered for the system GSC\,08047-00232\,A and B.}
\label{fig:cmd}
\end{figure}

We then used the evolutionary tracks of Baraffe et al. (1998) and Chabrier et al. (2000) as a comparison to test if the photometry of the faint candidate companions is consistent with that expected for very low-mass stellar or substellar companions. The evolutionary model of Baraffe et al. (1998) is a dust-free and non-gray atmosphere model (BCAH98) appropriate to describe M dwarfs. The DUSTY and COND models of Chabrier et al. (2000) can be seen as two extreme cases, to describe respectively brown dwarfs with an atmosphere saturated in dust, like late M and L dwarfs, and cool brown dwarfs or giant planets with dust condensed in their atmosphere, like T dwarfs. Based on the photometric results in Table~\ref{tab:photo}, the color-magnitude diagram is presented in Fig.~\ref{fig:cmd}. 
 
Based on the COND model predictions for an age of 10 and 50~Myr (no predictions available between these two ages), the two candidate companions to HIP\,6856 have photometry consistent with objects lighter than ten Jupiter masses. If bound,  the cc-1 and cc-2 would have respective physical separations of 180~AU and 104~AU. 

In the case of GSC\,08047-00232\,B, we considered a distance of of 85\,pc according to the JHKs photometry of GSC\,08047-00232\,A. This places GSC\,08047-00232\,B at a projected physical separation of 278~AU. Based on the DUSTY model predictions for an age of 10 and 50~Myr, GSC\,08047-00232\,B  has photometry consistent with a cool substellar companion with a mass of $25\pm15~\rm{M}_{\rm{Jup}}$ and an effective temperature of $2100\pm300$\,K. In addition, considering the relationship between photometry and spectral type of \citet{dahn02}, we derive a spectral type of L$0\pm2$ for this faint object.

%
%
%
\section{Proper Motions} 

\subsection{Two background objects close to HIP\,6856} 

Based on the astrometric measurements presented in Table~\ref{tab:astro}, the differences of separations for the cc-1 to HIP\,6856 between 26 November 2002 and the two successive epochs are reported in Fig.~\ref{fig:sep1}a. The expected variations (with their uncertainties) in the case of a background object are also given according to the proper motion of HIP\,6856 given by the Tycho catalog \citep{hog00}: $\mu_{\alpha}=104.8\pm1.2$~mas/yr and $\mu_{\delta}=-43.4\pm1.4$~mas/yr. We also considered the largest orbital variations expected in the case of a bound companion, based on Kepler's third law. The differences in terms of position angles are not considered as they are negligible in comparison to the differences in separations. Fig.~\ref{fig:sep1}a shows that the cc-1 is clearly not co-moving with HIP\,6856 with a divergence of 4.3\,$\sigma$ of being a bound companion.

Similarly, we report the differences of separations for the newly discovered cc-2 to HIP\,6856 (see Fig.~\ref{fig:sep1}b). This object is also likely a background object, with a discrepancy in separation of 4\,$\sigma$ of being bound. 
\subsection{The co-moving system GSC\,08047-00232 A and B}

\begin{figure}[t]
\centering
\includegraphics[width=7.63cm]{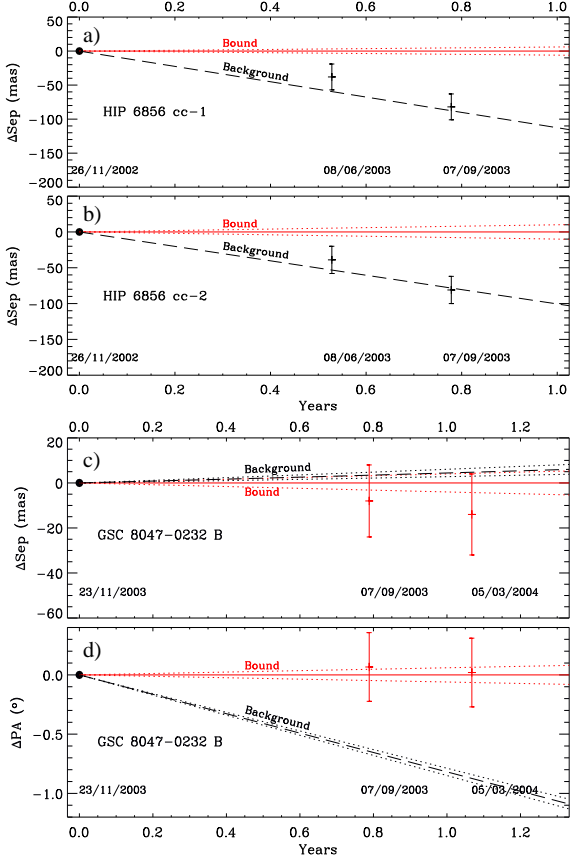}
\caption{\textbf{a)} Differences of separations, measured with NACO, for the cc-1 to HIP\,6856 between 26/11/2002 and the successive epochs 08/06/2003 and 07/09/2003. Expected evolutions with their uncertainties in case of bound (\textit{solid lines}) or background (\textit{dashed lines}) objects are given. \textbf{b)} idem but for the cc-2 to HIP\,6856. \textbf{c)} Differences of separations, measured with NACO, for GSC\,08047-00232\,B between 23/11/2002 and the two successive epochs 07/09/2003 and 05/03/2004. The expected evolutions  with their uncertainties for bound (\textit{solid line}) and background (\textit{dashed line}) objects are indicated. \textbf{d)} idem but for the differences of position angles for GSC\,08047-00232\,B.}\label{fig:sep1}
\label{fig:sep}
\end{figure}

According to the astrometric measurements of GSC\,08047-00232 A and B (see Table~\ref{tab:astro}), the difference in terms of separations and position angles between 23 November 2002 and the two successive epochs are reported in  Fig.~\ref{fig:sep1}c and Fig.~\ref{fig:sep1}d. We used the proper motion of GSC\,08047-00232\,A, $\mu_{\alpha}=46.9\pm1.7$~mas/yr and $\mu_{\delta}=-3.1\pm1.7$~mas/yr \citep{hog00}, to report also the expected variations (with their uncertainties) in the case of a background object. 
Contrary to the two non-true companions to HIP\,6856, GSC\,08047-00232\,B shares common proper motion with GSC\,08047-00232 A. The difference of position angles between 23 November 2002 and 5 March 2004 shows at the 3.1\,$\sigma$ confidence level that GSC\,08047-00232\,B is not a stationary background object.

%
%
%
\section{Spectral characterization of GSC\,08047-00232\,B} 

\begin{figure}[t]
\centering
\includegraphics[width=8.5cm]{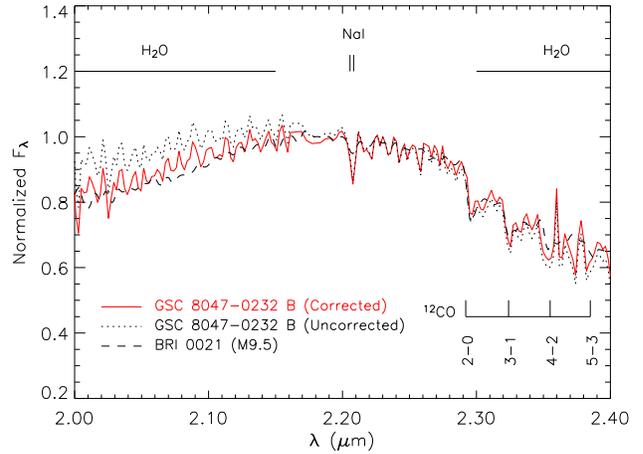}
\caption{SK-band spectra of GSC\,08047-00232\,B with the medium resolution ($\rm{R}=1400$) grism of CONICA, the 86~mas slit and the S54 camera (54~mas/pixel). The continuum slope of this spectrum has been increased by 14\% between 2.00 and 2.48~$\mu$m to best-match the template spectrum of  BRI\,0021 (M9.5). The spectra here have been normalized to the flux at 2.18~$\mu$m.}
\label{fig:spectrum}
\end{figure}
To finally confirm the companionship of GSC\,08047-00232\,B as well as its substellar nature, we recorded spectra on 25 November 2002 for comparison with the observations of M, late-M and L dwarfs published by \citet{reid01}, \citet{geba02} and \citet{legg01,legg00}. None of these template spectra matched well the continua of the GSC\,08047-00232\,B spectrum which appeared to be too steep. As already evoked by \citet{goto02}, the main plausible explanation seems to be a chromatical dependency of the flux transmitted through the narrow slit in case of a difference of centering between GSC\,08047-00232\,B and the standard star. This results in a decrease of the continuum slope going from short to long wavelengths, well represented by a linear function which is effectively observed in the case the GSC\,08047-00232\,B spectrum.

To correct this effect, we multiplied the GSC\,08047-00232\,B spectrum by a linear function where the slope has been adjusted to match the continuum of the template spectra. The best adjustement was obtained with the template spectrum of BRI\,0021 (M9.5) by increasing the continuum slope of 14\% between 2.00 and 2.48~$\mu$m (see Fig.~\ref{fig:spectrum}). We finally derived a spectral type M$9.5\pm1$ for GSC\,08047-00232\,B based on the other template spectra. To test if such variation could be explained by the chromatical effect described previously, we simulated the flux transmitted through the 86~mas (1.6 pixels) for several NACO PSFs\footnote{Simulations obtained with the \textit{NACO Preparational Software}, see http://www.eso.org/observing/etc/naosps-p71/doc/index.html.}. We considered different wavelengths (2.00, 2.18, 2.30 and 2.48~$\mu$m) and the seeing conditions of the GSC\,08047-00232\,B and HIP\,9022 observations (see Table~\ref{tab:setup}). We finally demonstrated a variation of the continuum slope of 14\% between 2.00 and 2.48~$\mu$m could be simulated with an offset of 1.2 to 1.5 pixels of the GSC\,08047-00232\,B spectrum from the centred position behind the slit. 

In addition to the continuum adjustment of GSC\,08047-00232\,B, the presence of deep absorption lines of NaI~(2.209\,$\mu$m) and the $^{12}$CO transitions 2-0 (2.295\,$\mu$m), 3-1 (2.324\,$\mu$m), 4-2 (2.354\,$\mu$m) and 5-3 (2.385\,$\mu$m) support the fact that GSC 08047-00232 B is a late-M dwarf. Using the $\textit{R}_{\rm{CO}}$ index of Reid et al. (2001), defined as the ratio between the flux at the base of the primary band head (at 2.29\,$\mu$m) and the pseudocontinuum flux at 2.27\,$\mu$m, we find a ratio of 0.83 consistent with spectral type M8.5-M9.

Our spectral type is consistent with that of Neuh\"auser \& Guenther (2004) who deduced M$8\pm2$ for GSC\,08047-00232\,B based on H- and K- band spectra respectively obtained with VLT/ISAAC and the SOFI spectrograph at the ESO/NTT.

According to \citet{legg01}, the range of spectral types derived above is consistent with T$_{eff}=2100\pm200$~K. This is consistent with the estimation based on the comparison between photometry and the DUSTY model predictions. Consequently, we confirm that GSC\,08047-00232\,B is a young brown dwarf companion to the star GSC 08047-00232 A, with a physical separation of 278~AU for a distance of 85~pc. This substellar companion is presently the third one detected among the young nearby associations with TWA5~B in the TW Hydrae association, $20~\rm{M}_{\rm{Jup}}$ at $\sim100$\,AU for $\textit{d}=50~$pc \citep{lowr99,neuh00}, and HR\,7329\,B  in the $\beta$ Pictoris Group, $20-30~\rm{M}_{\rm{Jup}}$ at $\sim200\,$AU for  $\textit{d}=47~$pc \citep{lowr00,guen01}.
 
%
%
%
\section{Conclusion} 

We present new VLT/NACO high contrast and high angular resolution observations of the two probable members of the Tucana-Horologium association: HIP\,6856 and GSC\,08047-00232. A faint candidate companion had been previously detected with ADONIS/SHARPII in the circumstellar environment of both stars \citep{chau03}. Determining companionship through proper motion and spectroscopy was the main purpose of this work. 

The enhanced detection capabilities of NACO allowed us to detect a second faint object close to HIP\,6856. We then checked that the photometry of these faint objects was consistent with that predicted by evolutionary model predictions for low-mass and cool objects. The two faint objects close to HIP\,6856 could be consistent with fewer than ten Jupiter masses bodies and  GSC\,08047-00232\,B with a brown dwarf companion.

Based on observations obtained at different epochs, we determined the proper motions of these faint objects and showed that the two candidate companions to HIP\,6856 are likely to be background objects. On the contrary, we confirm that GSC\,08047-00232\,B shares common proper motion with GSC\,08047-00232\,A. We finally corroborate the companionship of GSC\,08047-00232\,B based on the analysis of medium resolution spectra in the SK band. With spectral type M$9.5\pm1$, GSC\,08047-00232\,B has an effective temperature of T$_{eff}=2100\pm200$~K. According to the Chabrier et al. (2000) DUSTY evolutionary model for an age of 30~Myr, we derive a mass of $25\pm10~\rm{M}_{\rm{Jup}}$, confirming the substellar nature of this faint companion, the third one actually identified among young, nearby associations.  

\begin{acknowledgements}
We would like to thank the staff of the ESO Paranal observatory and the NAOS/CONICA team for their help in carrying out this work. We thank also Gilles Chabrier, Isabelle Baraffe and France Allard for providing us with the latest update of their evolutionary models. We thank also Sandy Leggett, Tom Geballe and Neill Reid who kindly sent us the near-infrared template spectra of M, late-M and L dwarfs for direct comparison with our measurements. Finally, we would like to thank Michael Sterzik for the different discussions about the spectroscopic analysis of AO observations. 
\end{acknowledgements}


\begin{thebibliography}{}
%
\bibitem[Baraffe et al.(1998)]{bara98} Baraffe I., Chabrier G., Allard F. \&
Hauschildt P.H., 1998, A\&A 337, 403
%
\bibitem[Becklin \&\ Zuckerman(1988)]{beck88} Becklin, E. \& Zuckerman, B. 1988, Nature, 336, 656
%
\bibitem[Chabrier et al.(2000)]{chab00} Chabrier, G., Baraffe, I., Allard, F. \& Hauschildt, P.H. 2000, ApJ, 542, 464
%
\bibitem[Chauvin et al.(2003)]{chau03} Chauvin, G., Thomson, M., Dumas, C. et al. 2003, A\&A, 406, 51
%
\bibitem[Cutri et al.(2003)]{cutr03} Cutri, R. M., Skrutskie, M. F., van Dyk, S. et al. 2003, 2MASS All-Sky Catalog of Point Sources
%
\bibitem[Devillar(1997)]{devi97} Devillar N. 1997, The messenger, 87
%
\bibitem[Dahn et al.(2002)]{dahn02} Dahn C.C., Harris H.C., Vrba F.J. et al. 2002, AJ, 124, 1170  
%
\bibitem[Elmegreen(1999)]{elme99} Elmegreen, B. G. 1999, ApJ 522, 915
%
\bibitem[Geballe et al.(2002)]{geba02} Geballe, T.R., Knapp, G.R., Leggett\, S.K. et al. 2002, ApJ, 564, 466
%
\bibitem[Gizis et al.(2001)]{gizi01} Gizis, J. E., Kirkpatrick, J. D., Burgasser, A. et al. 2001, ApJ 551, 163
%
\bibitem[Goto et al.(2002)]{goto02} Goto, M., Kobayashi, N., Terada H. et al. 2002, ApJ, 567, L59    
%
\bibitem[Guenther et al.(2001)]{guen01}Guenther, E. W., Neuh\"auser, R., Hu\'elamo, N., Brandner, W. \& Alves, J. 2001, A\&A, 365, 514
%
\bibitem[Halbwachs et al.(2000)]{halb00} Halbwachs, J. L., Arenou, F., Mayor, M., Udry, S. \& Queloz, D. 2000, A\&A 355, 581
%
\bibitem[H{\o}g et al.(2000)]{hog00}H{\o}g, E., Fabricius, C., Makarov, V.V. et al. 2000, A\&A, 355, 27
%
\bibitem[Leggett et al.(2000)]{legg00} Leggett, S.K., Geballe, T.R., Fan, X. et al. 2000, ApJ, 536, L35 
%
\bibitem[Leggett et al.(2001)]{legg01} Leggett, S.K., Allard, F., Geballe, T.R., Hauschildt, P.H. \& Schweitzer, A. 2001, ApJ, 548, 908 
%
\bibitem[Lenzen et al.(1998)]{lenz98} Lenzen, R., Hofmann, R., Bizenberger, P. \& Tusche, A., 1998, SPIE, Vol. 3354
%
\bibitem[Lowrance et al.(1999)]{lowr99} Lowrance, P. J., McCarthy, C., Becklin, E. E. et al. 1999, ApJ, 512, L69 
%
\bibitem[Lowrance et al.(2000)]{lowr00} Lowrance, P. J., Schneider, G,, Kirkpatrick, J. et al. 2000, ApJ, 541, L390
%
\bibitem[Lucas \&\ Roche(2000)]{lucas00} Lucas, P.W. \& Roche, P.F. 2000, MNRAS, 314, 858
%
\bibitem[Luhman et al.(2000)]{luhm00} Luhman, K., Rieke, G.H., Young, E.T. et al. 2000, ApJ, 540, 1016
%
\bibitem[McCarthy \& Zuckerman(2004)]{mcar04} McCarthy, C. \& Zuckerman, B. 2004, AJ, 127, 2871
%
\bibitem[Mizuno(1980)]{mizu80}Mizuno, H. 1980, Prog. Theo. Phys., 64, 544 
%
\bibitem[Moraux et al.(2003)]{mora03}Moraux, E., Bouvier, J., Stauffer, J.R. \& Cuillandre, J.-C. 2003, A\&A, 400, 891  
%
\bibitem[Nakajima et al.(1995)]{naka95} Nakajima, T., Oppenheimer, B.R., Kulkarni, S.R. et al. 1995, Nature, 378, 463
%
\bibitem[Neuh\"auser et al.(2000)]{neuh00}Neuh\"auser, R., Guenther, E. W., Petr, M. G. et al. 2000, A\&A, 360, 39
%
\bibitem[Neuh\"auser et al.(2003)]{neuh03} Neuh\"auser, R., Guenther, E.W., Alves, J. et al. 2003, AN, 324, 535
%
\bibitem[Neuh\"auser \& Guenther(2004)]{neuh04} Neuh\"auser, R. \& Guenther, E.W. 2004, A\&A, 420, 647
%
\bibitem[Pickett(2000)]{pick00}Pickett, B. K., Durisen, R. H., Cassen, P. \& Mejia, A. C. 2000, ApJ, 540, 95
%
\bibitem[Reid et al.(2001)]{reid01} Reid, I.N., Burgasser, A.J., Cruz, K.L., Kirkpatrick, J.D. \& Gizis, J.E. 2001, AJ, 121, 1710 
%
\bibitem[Reipurth \&\ Clarke(2001)]{reith01} Reipurth, B. \& Clarke, C. 2001, AJ, 122, 432
%
\bibitem[Rousset et al.(2002)]{rous02} Rousset, G., Lacombe, F., Puget, et al., 2002, SPIE, Vol. 4007
%
\bibitem[Torres et al.(2000)]{torr00} Torres, C.A.O., Da Silva, L., Quast, G.R., de la Reza, R. \& Jilinski, E. 2000, AJ, 120, 1410
%
\bibitem[Van der Bliek et al.(1996)]{vand96} Van der Bliek, N.S., Manfroid J. \& Bouchet, P. 1996, A\&AS, 119, 547
%
\bibitem[V\'eran \& Rigaut(1998)]{vera98} V\'eran, J.P. \& Rigaut, F. 1998, SPIE, 3353, 426 
%
\bibitem[Zuckerman \& Webb(2000)]{zuck00}Zuckerman, B. \& Webb, R.A. 2000, ApJ, 535, 959
%
\bibitem[Zuckerman \& Song(2004)]{zuck04}Zuckerman, B. \& Song, I. 2004, ARAA, 42, 685
\end{thebibliography}
\end{document}